\documentclass[aip,jcp,reprint,floatfix]{revtex4-1} 
\usepackage{mathtools}
\usepackage{amsmath}
\usepackage{multirow}
\usepackage[english]{babel}
\usepackage{amsmath,lipsum}
\usepackage{etoolbox}
\appto{\normalsize}{\zerodisplayskips}
\appto{\small}{\zerodisplayskips}
\appto{\footnotesize}{\zerodisplayskips}
\usepackage{textcomp}
\usepackage{gensymb}
\newcommand{\zerodisplayskips}{%
  \setlength{\abovedisplayskip}{5pt}%
  \setlength{\belowdisplayskip}{5pt}%
  \setlength{\abovedisplayshortskip}{5pt}%
  \setlength{\belowdisplayshortskip}{5pt} }
\appto{\normalsize}{\zerodisplayskips}
\appto{\small}{\zerodisplayskips}
\appto{\footnotesize}{\zerodisplayskips}
\usepackage{graphicx}
\usepackage{color}
\usepackage{enumitem}
\usepackage[dvipsnames]{xcolor}
\usepackage[normalem]{ulem}

\renewcommand\thesubsection{\arabic{subsection}}
\renewcommand\thesubsubsection{\arabic{subsubsection}}
\makeatletter
    \def\@seccntformat#1{\@ifundefined{#1@cntformat}%
       {\csname the#1\endcsname\space}%    default
       {\csname #1@cntformat\endcsname}}%  enable individual control
    \def\subsection@cntformat{\thesection.\thesubsection\space} 
    \def\subsubsection@cntformat{\thesection.\thesubsection.\thesubsubsection\space}
\makeatother

\DeclareUnicodeCharacter{0301}{\'{e}} % Foreign accent
\DeclareUnicodeCharacter{2009}{\,} % Foreign accent

\begin{document}

\title{Towards automated sampling of polymorph nucleation and free energies with SGOOP and metadynamics}

\author{Ziyue Zou}
\affiliation{Department of Chemistry and Biochemistry, University of Maryland, College Park 20742, USA.}
 
\author{Sun-Ting Tsai}
\affiliation{Department of Physics, University of Maryland, College Park 20742, USA.}
\affiliation{Institute for Physical Science and Technology, University of Maryland, College Park 20742, USA.}

\author{Pratyush Tiwary*\thanks{ptiwary@umd.edu}}
\email{ptiwary@umd.edu}
\affiliation{Department of Chemistry and Biochemistry, University of Maryland, College Park 20742, USA.}
\affiliation{Institute for Physical Science and Technology, University of Maryland, College Park 20742, USA.}

\date{\today}

\begin{abstract}
Understanding the driving forces behind the nucleation of different polymorphs is of great importance for material sciences and the pharmaceutical industry. This includes understanding the reaction coordinate that governs the nucleation process as well as correctly calculating the relative free energies of different polymorphs. Here we demonstrate, for the prototypical case of urea nucleation from melt, how one can learn such a 1-dimensional reaction coordinate as a function of pre-specified order parameters, and use it to perform efficient biased all-atom molecular dynamics simulations. The reaction coordinate is learnt as a function of generic thermodynamic and structural order parameters using the ``Spectral Gap Optimization of Order Parameters (SGOOP)" approach [P. Tiwary and B. J. Berne, Proc. Natl. Acad. Sci. (2016)], and is biased using well-tempered metadynamics simulations. The reaction coordinate gives insight into the role played by different structural and thermodynamics order parameters, and the biased simulations obtain accurate relative free energies for different polymorphs. This includes accurate prediction of the approximate pressure at which urea undergoes a phase transition and one of the metastable polymorphs becomes the most stable conformation. We believe the ideas demonstrated in thus work will facilitate efficient sampling of nucleation in complex, generic systems.

\end{abstract}

\maketitle

\section{Introduction}
\label{sec:introduction}

Crystal nucleation is known to be the first step of crystallization, a process ubiquitous across biological and chemical systems. Understanding how different polymorphs can form has practical importance in material science \cite{Mon2014PolyinMat,Gen2019PolyinMat} and the pharmaceutical industry.\cite{Llinas2008polymorpha, brittain2018polymorphism, Lu2010polymorphism} However, experimental measurements suffer resolution problems, as nucleation events involve the sub-nanometer lengthscale and sub-picosecond timescale.\cite{Kal2016GrowthDesign} In principle, computational methods such as Molecular Dynamics (MD) simulations can provide detailed information about nucleation process since by construction they follow all-atom and femtosecond resolution. Unfortunately, this femtosecond temporal resolution limits the time scale that MD simulation can reach to around milliseconds, preventing it from sampling nucleation events which typically occur at seconds and much slower timescales.\cite{sosso2016crystal} Moreover, liquid-solid nucleation is further complicated by the presence of multiple competing crystal structures or complicated polymorphism, \cite{Bern2014poly} which is shown to be important in the pharmaceutical industry, material science and elsewhere.\cite{Mon2014PolyinMat,Gen2019PolyinMat, Llinas2008polymorpha, brittain2018polymorphism, Lu2010polymorphism} Understanding the nucleation of these polymorphs is therefore essential since different polymorphs lead to different physical and chemical properties.\cite{sosso2016crystal} 

In order to overcome the timescale limitation faced by MD simulations in the study of nucleation and more generally, several enhanced sampling methods have been developed.\cite{laio2002escaping,Torrie1977US,PhysRevLett2005FFS,JCP2006FFS,rosales2020seeding,bazterra2002modified} A large fraction of these methods needs a prior estimate of a biasing or progress coordinate that can differentiate relevant metastable states from each other, and can also quantify the progress of the slow process being studied. In principle, many approximate choices of this coordinate are sufficient for reliably enhanced sampling. In practice however, the enhancement in sampling is tremendously benefited if the biasing or progress coordinate approximates the true reaction coordinate (RC) for the process being studied. For the study of nucleation, the first such coordinate one can think of comes from classical nucleation theory (CNT)\cite{gibbs1928collected} which posits the size of a nucleus assumed spherical as a progress coordinate. CNT can however give wrong nucleation rates even for liquid Argon that can be off by almost twenty orders of magnitude. \cite{tsai2019reaction} This has led to a large effort over the decades aimed at developing better variables for quantifying the onset of nucleation. \cite{Schenter1999DNT, sosso2016crystal,Sarupria2017nucleation, Parrinello2021cv, Tribello2017analyzing, Trout2011orientationop,desgranges2018crystal} At the same time, a large number of techniques have been developed for the more general problem of learning approximate reaction coordinates for rare event systems given limited sampling. \cite{baron2006obtaining, Preto2014diffusionmapMD, tiwary2016spectral, pande2017TicaMeta,  Chiavazzo2017altas, mardt2018vampnets, ribeiro2018reweighted, Shamsi2018RLbasedAS, wang2019past} The problem of constructing good biasing or progress coordinates is especially accentuated when it comes to studying polymorphism  because such crystal structures all fall into a narrow range of energies \cite{bazterra2002modified} and very likely have similar orientations. 

In this work, we demonstrate how one such method ``Spectral gap optimization of order parameters (SGOOP)" \cite{tiwary2016spectral} can be used in conjunction with well-tempered metadynamics \cite{PRL2008WTMetaD} for sampling different polymorphs and accurately capturing their relative free energies. While the methods are described in greater detail in Sec.~\ref{sec:sgoop}, here we give an overview. Metadynamics involves constructing a bias potential by depositing history-dependent Gaussian kernels along the chosen biasing coordinate, which is typically low-dimensional, to encourage state-to-state transitions. SGOOP helps construct such a biasing coordinate as a linear combination of several order parameters (OPs). The central idea behind SGOOP is that the optimal RC will maximize the spectral gap of the dynamics projected along it, and the spectral gap itself can be learnt efficiently from biased metadynamics runs along trial coordinates, through a maximum path entropy or caliber framework. \cite{tiwary2016spectral, ghosh2020maximum}

In order to construct the RC as a combination thereof, we have many possible OPs at our disposal.\cite{Parrinello2021cv} Broadly speaking these fall into two classes.
The first type of OPs require pre-knowledge of the target crystal structure, and are based upon comparing the instantaneous configuration of a system with a specific reference structure. These include the Steinhardt bond order parameters\cite{steinhardt1983op} that compare systems with different spherical harmonics. Another example is the SMAC order parameter introduced by Giberti et al.\cite{SMAC2015insight} which describes the systems by selected angles between feature vectors. The second type of OPs are more generic and involve observables such as mean distance, mean dihedrals, and thermodynamics variables such as entropy and enthalpy. The first class of OPs is not very well suited for the study of polymorphism because of the loss in generality when specifying orientations. Thus in this work, we mainly focus on the second type and the detailed definitions are given in Sec. \ref{sec:op}. An optimized RC is constructed  as a function of those order parameters through SGOOP.\cite{tiwary2016spectral}

We illustrate the framework by studying urea polymorph nucleation from the melt. The urea system has been widely studied both experimentally\cite{bridgman1916polymorphism,Weber2002urea,Lamelas2005ureaXray,kat2009urea,donnelly2015urea,bini2017urea,roszak2017urea,safari2021highpressure} and computationally.\cite{Cheng2017ureaPES,SMAC2015insight,piaggi2018predicting} For consistency, we used the same notations of polymorphs of urea as in Ref.~\onlinecite{Cheng2017ureaPES}. In particular, form I and IV have been synthesized in lab and studied in detailed, whereas form A and B have been found only in simulations so far. \cite{Cheng2017ureaPES,SMAC2015insight,piaggi2018predicting} Ref.~\onlinecite{piaggi2018predicting} suggests that order parameters such as orientational entropy play an important role in urea nucleation and in exploring the configuration space relevant to the processes at work. We apply SGOOP to build a 1-d RC as a function of these entropy variables as well as other OPs that we describe in Sec.\ref{sec:op}. By performing well-tempered metadynamics along this optimized 1-d RC,  we obtain correct ranking of thermodynamic stabilities of different polymorphs. The use of such an optimized 1-d RC for enhanced sampling lowers the computational cost relative to non-optimized or higher-dimensional biasing variables. Finally, by extrapolating on the basis of the free energy difference obtained from our simulations, we obtain good agreement with experiments regarding the pressure at which one of the metastable polymorph becomes the more stable form.

%However this work was unable to use all-atom MD to obtain correct rankings of the polymorphs and their free energies relative to the melt form (!!check my claim!!) 

\begin{figure*}
  \centering
  \includegraphics[width=12cm]{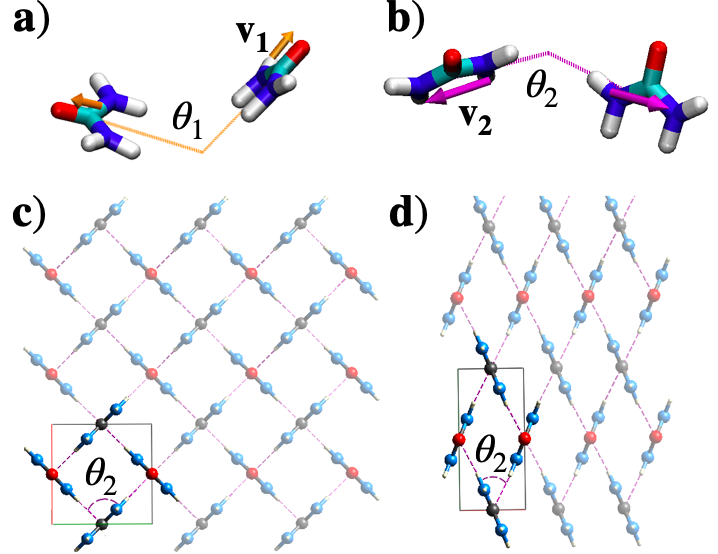}
  \caption
  {Definition of characteristic vectors, $\bf{v}_1$ and $\bf{v}_2$ in the upper panel (a and b, respectively). Configurations of forms I ($P\bar{4}2_1m$) and IV ($P2_12_12$) with $\bar\theta_2$ angle along [001] in the lower panel (c and d). $\bf{v}_1$ is defined as the dipole moment of urea molecules and $\bf{v}_2$ connects the two nitrogen atoms. Direct measure of $\bar\theta_2$ order parameter allows us to distinguish these two forms. Crystallographic data of forms I and IV were obtained from Cambridge Structural Database (CCDC 731958, 731961).\cite{kat2009urea} Structures were gerenated with Visual Molecular Dynamics (VMD) \cite{HUMPHREY1996VMD} and CrystalExplorer 17.5.\cite{spackman2021CrystalExplorer}
  }
  \label{fig:thetaop}
\end{figure*}

\section{Methods}
\label{sec:method}

\subsection{Order parameters for nucleation}
\label{sec:op}
Many order parameters have been developed to study the nucleation process.\cite{Sarupria2017nucleation, Parrinello2021cv, Tribello2017analyzing, duff2011polymorph, Trout2011orientationop, fulford2019NNIce, Duboue2015waterOP, Doi2020Mining} Here we consider four types of order parameters. which are then used to construct an optimized 1-d RC as a function thereof. These are (i) pair orientational entropy, \cite{piaggi2018predicting,Piaggi2017entropy} (ii) enthalpy, (iii) averaged intermolecular angles, and (iv) mean coordination number. We now describe them systematically:

\begin{enumerate}[leftmargin=*]
\item Pair orientational entropy:
Piaggi et al.\cite{piaggi2018predicting} recently introduced an approximation to the pair entropy as a biasing variable for metadynamics. Referring to liquid state theory,\cite{green1958liquidstate} the excess entropy with respect to the entropy of ideal gas can be expressed in an infinite sum of correlation functions, and has been shown as a successful variable for analysis by considering only the first term.\cite{entropyWater1996JCP, entropyGlass2000PRE, entropyWater2010JCTC, entropyWater2011JPCB} The excess pair orientational entropy is then defined as:
\begin{align}
S_\theta = -\pi \rho k_{B} &\int_{0}^\infty \int_{0}^\pi [g(r,\theta)\ln{g(r,\theta)} \nonumber \\
&-g(r,\theta)+1]r^2\sin{\theta}drd\theta
  \label{eq:S_theta}
\end{align}
where $\rho$ is the density of the system, $k_{B}$ is Boltzmann constant, $r$ is the intermolecular distance, $\theta$ is the angle formed between characteristic vectors, and $g(r,\theta)$ is a mollified correlation function\cite{piaggi2018predicting} to ensure its differentiability for enhanced sampling methods. Given the polarity of urea molecules, two characteristic vectors are introduced (Fig.~\ref{fig:thetaop} upper panel.). The first vector \textbf{v}$_1$ is along the direction of the carbonyl group, which is also the direction of its dipole moment. The second vector \textbf{v}$_2$ connects the two nitrogen atoms. Angles that are formed between a pair of two \textbf{v}$_1$ vectors on two adjacent urea molecules are denoted  $\theta_1$  while those between \textbf{v}$_2$ are denoted as $\theta_2$. Using these two angles in Eq.~\ref{eq:S_theta}, we define two representations of entropy denoting them $S_{\theta_1}$ and $S_{\theta_2}$.

\item Enthalpy: To complete the picture of thermodynamics related variables, enthalpy was introduced as an order parameter in Ref.~\onlinecite{Piaggi2017entropy}. Unlike entropy which requires approximations to be calculated, enthalpy can be computed directly from its definition:
\begin{align}
H = E + PV
\label{eq:h}
\end{align}
where $E$ is the instantaneous total potential energy, $V$ is the volume and $P$ is the pressure. Given the simulation is performed at isothermal-isobaric ensemble, pressure is a constant. 

\item Mean coordination number: We consider the mean coordination number for urea molecules in light of several studies on gas/liquid nucleation \cite{barducci2011cn,matteo2016cn,tsai2019reaction} and we calculate this as \cite{barducci2011cn}
\begin{align}
&N = \frac{\sum_{ij}\frac{1-(r_{ij}/r_c)^6}{1-(r_{ij}/r_c)^{12}}}{n}
\label{eq:cn}
\end{align}
where n is the total number of molecules, $r_c$ is a cutoff and $j$ denotes molecules besides molecule $i$. For molecules with distance $r_{ij}$ within this threshold distance, they are identified as neighbors. Similarly, we defined the cutoff distance as the radius of first shell, 0.64 $nm$. While we do not expect that $N$ will be able to distinguish different polymorphs from each other, it is well-documented to be useful for differentiating liquid from solid phases and thus we consider it here as one of the possible constituents of the reaction coordinate.

\item Averaged intermolecular angles: To further distinguish polymorphs from each other, here we introduce another order parameter that is based on the averaged values of the angles $\theta_1$ and $\theta_2$ introduced earlier (Fig.~\ref{fig:thetaop} upper panel). For either of the two angles, this variable is defined as follows:
\begin{align}
&\bar\theta=\frac{\sum_{ij}\sigma(r_{ij})f(\theta_{ij})}{\sum_{ij}\sigma(r_{ij})}
\label{eq:cd_2RCs}
\end{align}
where $\sigma(r_{ij})$ is a smooth switching function of intermolecular distance $r_{ij}$, and $f(\theta) = \frac{1}{2}[(\pi-2\theta)\tanh(5\theta-9.25)+\pi]$ is a step function of angle $\theta$ in a form of hyperbolic tangent that ensures continuous differentiability. This converts $\theta$ to its supplementary angle when such angle is larger than a cutoff $\theta_{ij}$. Considering the $C_{2v}$ symmetry of urea molecule, permutation of \textbf{v}$_2$ is unavoidable by directly calculating average angles. One alternative is to define a step function to introduce rotational invariance as achieved through the function $f$.  This order parameter is developed for the purpose of differentiating form I and form IV along the direction of \textbf{v}$_2$, where form IV ($P2_12_12$) is known to be a high-pressure product of urea in experiment and form I ($\bar{P}42_1m$) is the most stable structure found at ambient conditions.\cite{Lamelas2005ureaXray,kat2009urea,bini2017urea,safari2021highpressure} The main difference of these two polymorphs are illustrated by $\theta_2$, as it equals 50$\degree$ for form IV and 90$\degree$ for form I (shown in Fig.~\ref{fig:thetaop} lower panel).\cite{kat2009urea} In this work, the threshold of the step function of measured angles is set to be 100$\degree$, which shows better performance than a cutoff of 90$\degree$ in distinguishing form I from its collapsed version, and a cutoff of distance is set to be 0.64 $nm$ corresponding to the range of first neighbor shell.

\end{enumerate}

\subsection{Spectral gap optimization of order parameters (SGOOP)}
 \label{sec:sgoop}
 
While the different order parameters (OPs) described in Sec.~\ref{sec:op} can collectively differentiate polymorphs from each other, it is not obvious how to use these OPs to perform enhanced sampling of polymorph nucleation. Considering all 6 OPs collectively as biasing or progress coordinates in a protocol such as metadynamics\cite{laio2002escaping}, umbrella sampling\cite{Torrie1977US} or forward flux sampling\cite{PhysRevLett2005FFS,JCP2006FFS,haji2015ffs,Sarupria2017nucleation, Sosso2016microscopic} is possible in principle, but not in practice.\cite{Defever2019cFFS} Thus, our next objective is to compress these OPs into a lower-dimensional approximate reaction coordinate (RC) which best encapsulates the relevant slow degrees of freedom. In this work, we use the method ``Spectral Gap Optimization of order parameters (SGOOP)" \cite{tiwary2016spectral} to find such an optimal 1-d RC as a linear combination of the OPs we defined in Sec. \ref{sec:op}. SGOOP utilizes the principle of Maximum caliber (MaxCal) \cite{presse2013principles, dixit2015inferring, ghosh2020maximum} to construct estimates of transition probability matrices along several trial RCs. This is done by combining stationary density information from biased simulations along trial RCs and dynamical information from short unbiased simulations. Ref. \onlinecite{tiwary2017predicting} illustrated through numerical examples how the RC learnt from SGOOP provides a very good approximation to those from more accurate but expensive methods. We now summarize the key concepts behind SGOOP. Let $\pi_m$ denote the stationary distribution along a spatially discretized trial RC with $m$ denoting the grid index. This can be obtained from reweighting a biased simulation performed along any preliminary RC approximation \cite{reweighting2015jpcb}. In addition to $\pi_m$, we also introduce $\langle N\rangle$ as the average number of nearest-neighbor transitions along any spatially discretized trial RC per unit time, which provides dynamical constraints to be used in the MaxCal framework. $\langle N\rangle$ can be estimated by directly counting from short unbiased MD trajectories. With these two pieces of stationary and dynamical information, MaxCal provides a tractable approximation for the transition matrix $K$ along any trial RC:
\begin{align}
    K_{mn}=-\frac{\langle N\rangle}{\sum\sqrt{\pi_m\pi_n}}\sqrt{\frac{\pi_n}{\pi_m}}
    \label{eq:MaxCal_K}
\end{align}
 For any putative RC, one can construct such a transition probability matrix and diagonalize it to get eigenvalues $\lambda_0=1>\lambda_1>\lambda_2>...>0$. These eigenvalues directly provide information about the timescales of different transition processes when projected on the corresponding RC along which the matrix was built. We can then find an optimal RC $\chi$ by selecting from the trial RCs the one with maximal spectral gap which is defined as the maximal difference between the visible slow and hidden fast processes. Such differences can be easily quantified as $\lambda_n-\lambda_{n+1}$ where $n$ is the number of visible slow processes that we really care about. Note that here we restrict ourselves to a 1-d RC as a trade-off between achieving accurate description of the slow processes while staying computationally efficient. For more complicated systems, recent developments using SGOOP allow us to systematically quantify if more dimensions need to be added to the RC.\cite{tsai2021sgoopd}

\subsection{Simulation setup}
\label{sec:simulation}
The system consists of 108 urea molecules described using generalized Amber force field (GAFF).\cite{amber_gaff} The partial charges of urea were adopted from Amber03 database \cite{amber03}. Biased and unbiased MD simulations were performed with a modified version of GROMACS-2020.2 \cite{abraham2015gromacs} patched with PLUMED 2.6.1 \cite{plumed2019nature,plumed2} in constant number, pressure and temperature (NPT) ensemble with an integration time step of 2 fs. The system was coupled with a thermostat using stochastic velocity rescaling method \cite{bussi2007canonical} at 450 K with 0.1 ps relaxation time. The pressure was controlled using the Parrinello-Rahman barostat \cite{parrinello1981polymorphic} at 1 bar with 10 ps relaxation time. The long-range electrostatic interaction in reciprocal space was calculated with Particle-mesh Ewald method.\cite{pme} The cutoffs for both electrostatic and Van der Waals interactions in real space were set to 0.9 nm. The hydrogen bonds were constrained with LINCS algorithm.\cite{lincs} We used well-tempered metadynamics (WTmetaD) \cite{barducci2008well, valsson2016enhancing} to enhance the sampling, where Gaussian kernels with width and height equal to 0.2 RC units and 7.48 $kJ/mol$ ($\approx 2 k_BT$) were deposited every 1 $ps$. The bias factor for the tempering was 50 and the production runs were performed for 400 $ns$.

\section{Results}
\label{sec:results}

\begin{table}[b]
    \centering
    \caption{Landmarks for stable configurations of urea. }
    \begin{tabular}{c|c|c|c|c|c|c}
    \hline \hline
          Phase & $S_{\theta_1}$ & $S_{\theta_2}$  & $\bar{\theta}_1$ & $\bar{\theta}_2$ & $N$ & $H$\\ 
          \hline 
         I & -6.11 & -2.12 & 0.257 & 1.22 & 11.609 & -766.93  \\
         \hline
         IV & -4.95 & -3.57 & 0.248 & 0.428 & 11.516 & -764.61\\ 
         \hline
         A$^*$ & -4.20 &  -5.79 & 0.389 & 0.411 & 12.288 & -763.25\\ 
         \hline
         B$^*$ & -1.76 & -3.39 & 1.272 & 0.666 & 11.543 & -762.99 \\
         \hline
         Liquid & -1.13 & -1.14 & 1.040 & 0.936 & 11.650 & -760.12 \\
         \hline \hline 
    \end{tabular}
    \newline $^*$Polymorphs not seen when biasing $\chi_6$ or in experiments but were seen in Ref.\onlinecite{piaggi2018predicting}.
    \label{tab:landmark}
\end{table}

\subsection{Landmarks for locating polymorphs}
\label{sec:landmarks}
In order to locate existing polymorphs during our enhanced sampling simulations, we first establish their corresponding landmarks in the order parameter space. The landmark for each polymorph can be found by the average of each of the 6 order parameters obtained in 2 $ns$ long unbiased simulation initiated from their individual configurations. These landmark values in CV space are then used as the fingerprints for tracing the formations of different polymorphs. Our primary interest in this work is in the polymorphs I and IV which have been synthesized in experiments. \cite{kat2009urea} These forms can be identified most cleanly simply using the two averaged angles $(\bar{\theta}_1, \bar\theta_2)$ introduced in Sec. \ref{sec:op}. We have also shown the landmark values for forms A and B in Table \ref{tab:landmark} even though they have not been seen in experiments yet, but were reported in past computational work. \cite{piaggi2018predicting} Forms A and B can be easily identified by  using $(S_{\theta_1},S_{\theta_2})$.

\begin{figure}
  \centering
  \includegraphics[width=8cm]{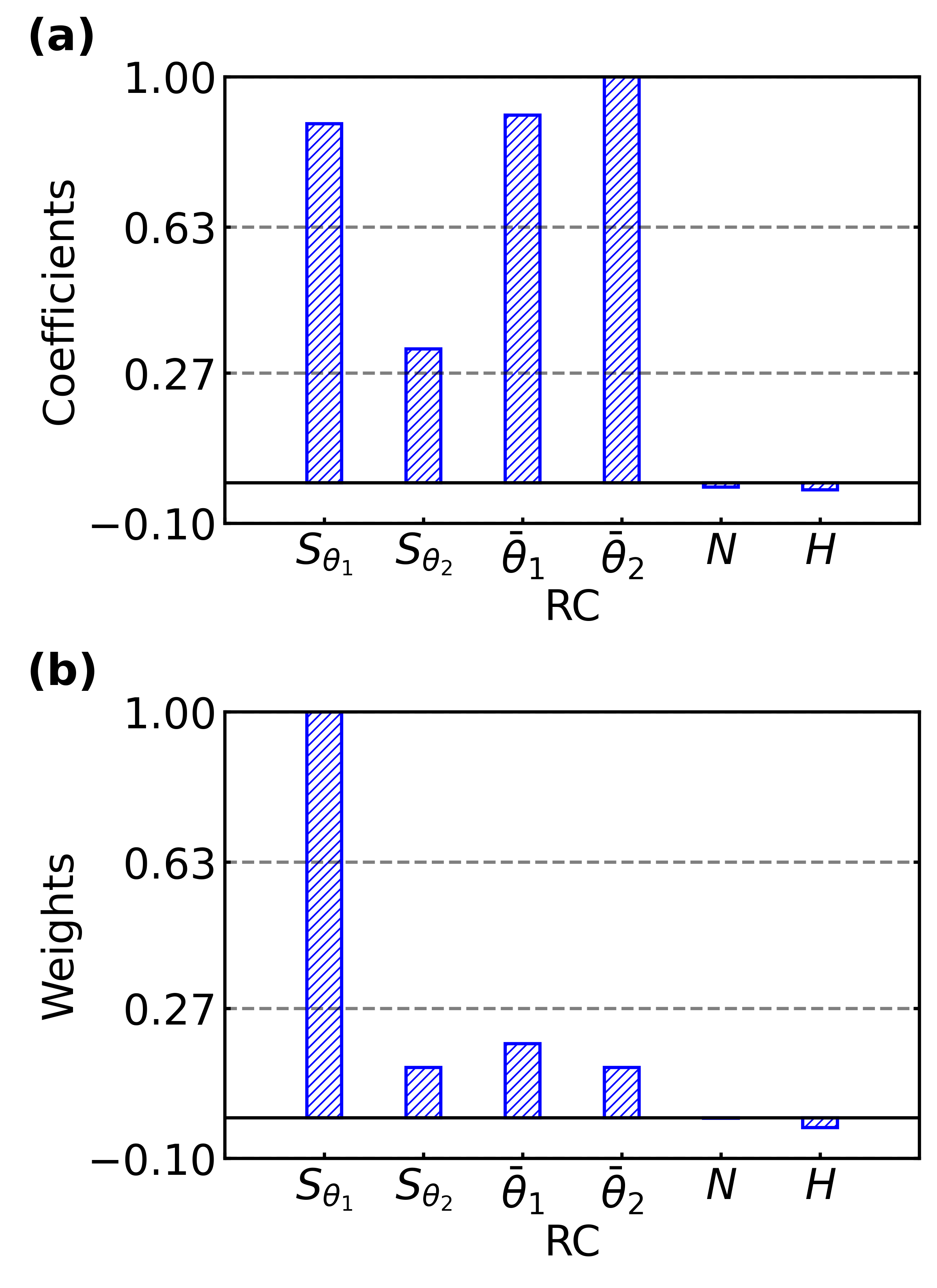}
  \caption
  {Bar plots of a) coefficients of 1-d reaction coordinate optimized by SGOOP and b) same coefficients, but scaled by their relative standard deviations. The weights in (b) are rescaled properly with respect to corresponding reweighted standard deviations observed in the different OPs of 9 individual 100 $ns$ long WTmetaD simulations for SGOOP optimization (details can be found in SI) and thus represent relative importance of each order parameters involved in the 1-d reaction coordinate. 
  }
  \label{fig:sgoopweights}
\end{figure}

\subsection{Reaction coordinate, sampling and converged free energies}
\label{sec:reac_coord}
By maximizing the spectral gap value using SGOOP (Sec. \ref{sec:sgoop}) using a basin-hopping algorithm \cite{Wales1997basinhopping}, we find the optimized RC as $\chi_6=0.885S_{\theta_1} +0.330S_{\theta_2} +0.906 \bar\theta_1 +1.0\bar\theta_2 -0.011N -0.017H$. See SI for further analysis of SGOOP results. By directly looking at the coefficients of the different OPs in this RC (Fig.~\ref{fig:sgoopweights} (a)), it can be seen that all 4 OPs except the mean coordination number $N$ and the enthalpy $H$ contribute to the RC. Yet another way to compare the contributions of different OPs is after accounting for their different scales, which can be done by dividing out their respective standard deviations from the coefficients in Fig.~\ref{fig:sgoopweights} (a). These standard deviations were computed from reweighting 9 individual 100 $ns$ long biased MD simulations along $S_{\theta_1}$ and $S_{\theta_2}$ starting from the melt, which are the same input trajectories for SGOOP algorithm (details given in SI). The subsequent weights are shown in Fig.~\ref{fig:sgoopweights}(b). Here we can see that $N$ and $H$ still have minimal contributions to the RC, but the pair orientational entropy $(S_{\theta_1}, S_{\theta_2})$, especially $S_{\theta_1}$, now has the most dominant weight. This is reminiscent of Argon liquid-droplet nucleation where second moment of coordination numbers have significantly higher weights. \cite{tsai2019reaction}

Interestingly, even though the weight for $H$ is small, the SGOOP RC $\chi_6$ captures a negative correlation between entropy and enthalpy in agreement with the definition  $G=H-TS$ of the free energy. The mean coordination number $N$ which is a measure of {number of} neighboring molecules, useful primarily for driving phase transitions from liquid to solid, does not necessarily have the ability to classify different polymorphs as solid urea are mainly face center cubic (FCC). As such it is not entirely surprising that it does not have a high weight in the SGOOP derived RC. In the SI we also report another RC found as a local maximum solution to the spectral gap optimization procedure but with lower spectral gap than the RC reported here. As shown in SI, the accuracy of the sampling performed using this sub-optimal RC is found to be poorer than the one shown here, reflecting the predictive power of the spectral gap metric of SGOOP. Finally, given the very low weights for $H$ and $N$ in $\chi_6$, we performed SGOOP to obtain a RC composed of only the other 4 OPs, namely $S_{\theta_1}$, $S_{\theta_2}$, $\bar\theta_1$, and $\bar\theta_2$. Enhanced sampling performed along this RC also shows slightly worse than with $\chi_6$ but still relatively decent convergence, consistent with our observation that $N$ and $H$ are relatively, though not entirely, unimportant for polymorph nucleation.

\begin{figure}
  \centering
  \includegraphics[width=8cm]{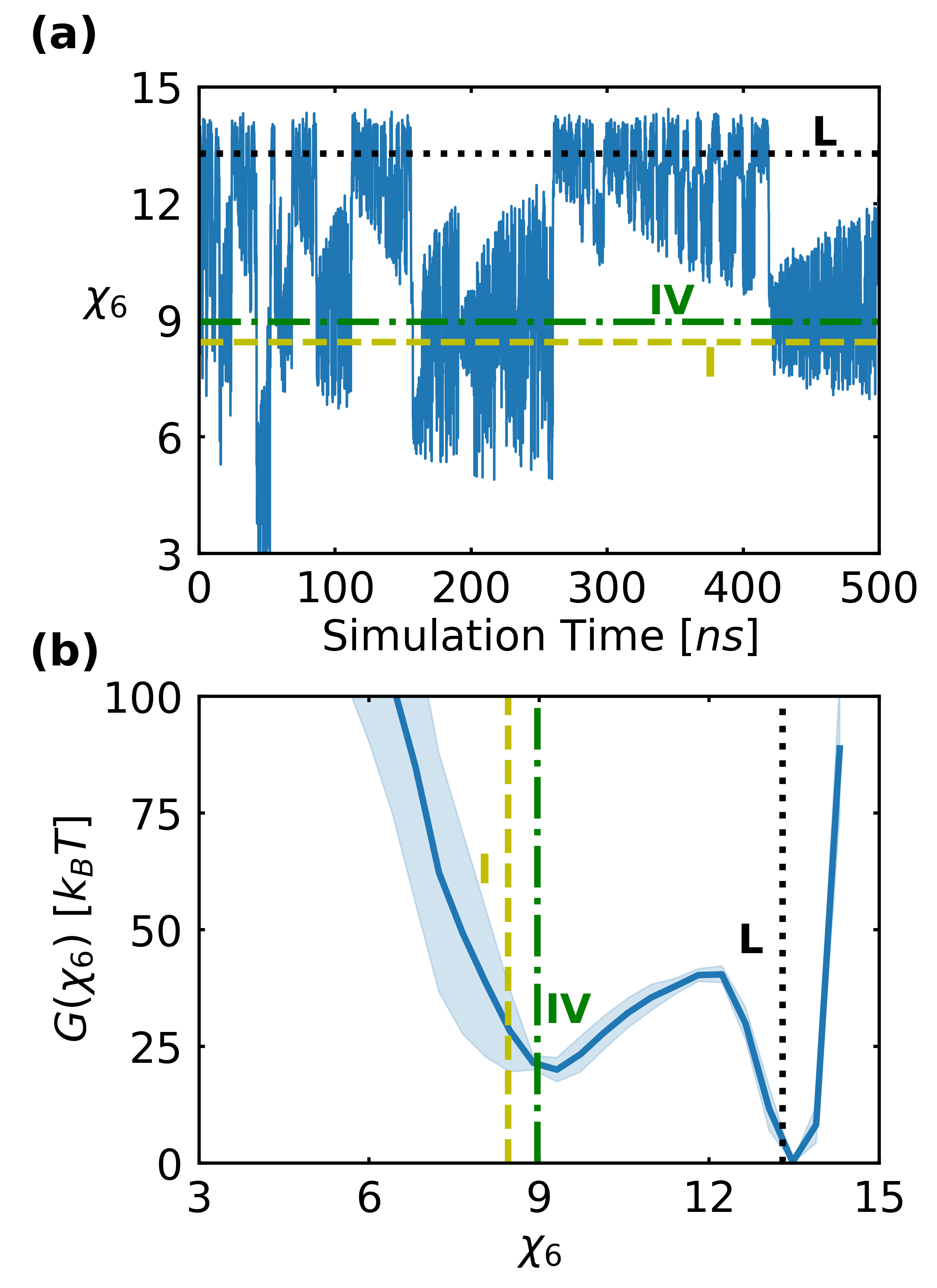}
  \caption
  { The time series of WTMetaD simulation biasing 1-d SGOOP optimized RC $\chi_6$ (a) and its associated reweighted free energy surface averaged over four individual 1-d WTmetaD simulations. (b) Forms I (in yellow) and IV (in green) were frequently visited in the trajectory and corresponding phases have been labeled in free energy profile. Error bars for the free energy are shown through shaded blue color. An unstable structure in which all dipoles are along the same direction is also sampled once throughout the simulation at 42 $ns$.
  }
  \label{fig:sgooprun}
\end{figure}

Once we have the optimized RC $\chi_6$ described in the previous two paragraphs, we perform well-tempered metadynamics (WTmetaD) simulation biasing this RC starting from liquid state.
The time series of the RC $\chi_6$ obtained from WTmetad is shown in Fig.\ref{fig:sgooprun}(a), where frequent state-to-state transitions occur and no evidence of substantial hysteresis, i.e. system getting trapped in a state is found suggesting $\chi_6$ as a good approximation to the RC. 
Fig.~\ref{fig:sgooprun}(b) shows the free energy profile along $\chi_6$ obtained from averaging over four independent WTmetaD simulations.  Fig.\ref{fig:sgooprun}(b) shows a distinctive barrier of height 45 $k_BT$ (i.e. 168 $kJ/mol$) between liquid (black dotted line in Fig.~\ref{fig:sgooprun}(a)) and solid states. It is interesting to note that this 1-d profile along $\chi_6$ does not show a discernible barrier between forms I and IV, even though our simulations did not get trapped in these states. However this is evidence that even for urea nucleation in vacuum one should be able to improve the RC further most likely by considering further order parameters that could contribute to the RC.

\begin{figure}
  \centering
  \includegraphics[width=8cm]{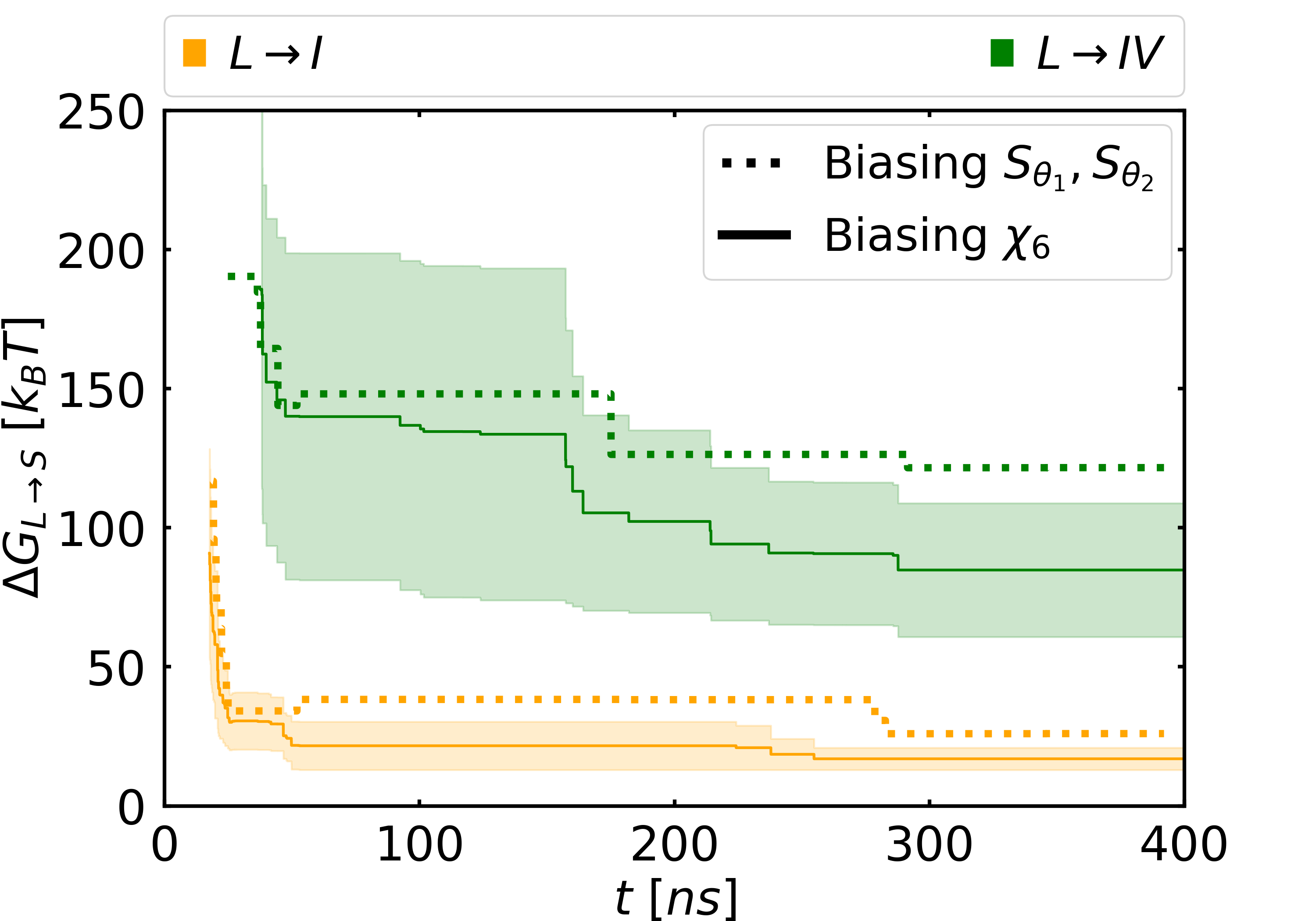}
  \caption
  {The free energy difference $\Delta G$ of forms I (in orange) and IV (in green) with respect to liquid phase over time from 2-d WTmetaD simulation biasing $S_{\theta_1}$ and $S_{\theta_1}$ (dotted lines) and 1-d WTmetaD biasing $\chi_6$ (solid lines). A 400 $ns$ trajectory of 2-d WTmetaD was reproduced from Ref.~\onlinecite{piaggi2018predicting}. Four independent simulations were collected and averaged over for the computation of free energy differences and error bars are shaded with corresponding colors. Form I has much lower free energy than that of form IV from both simulations and this agrees with experimental observations.\cite{kat2009urea}}
  \label{fig:freeEdiff}
\end{figure}

\subsection{Comparison with experimental measurements}
\label{sec:compare_experiments}
We now evaluate the quality of our RC and of the sampling so obtained by comparing with experimental observations. While it is difficult to make quantitative comparisons to experiments due to finite-size effects,\cite{matteo2016cn,salvalaglio2015molecular} we expect that the stable polymorph will have lower free energy than the unstable polymorphs. Here we define the free energy difference between any 2 phases $\alpha$ and $\beta$ as
\begin{align}
    \Delta G_{\alpha \rightarrow \beta} \equiv -G_\alpha + G_\beta \equiv \frac{1}{\beta}\log{\frac{P_\alpha}{P_\beta}} 
    \label{eq:FreeEDiff}
\end{align}
where $\beta$ is the inverse temperture and $p_\alpha$ and $p_\beta$ are the probabilities of the system being in phase $\alpha$ and $\beta$ respectively. We obtain these probabilities by reweighting the biased metadynamics simulations.\cite{reweighting2015jpcb}  Fig. \ref{fig:freeEdiff} shows the free energy differences for the different experimentally relevant phases through the use of  Eq. \ref{eq:FreeEDiff} by averaging over four independent WTmetaD simulations biasing $\chi_6$. Fig. \ref{fig:freeEdiff} also shows the results from 2-dimensional  WTmetaD biasing both $S_{\theta_1}$ and $S_{\theta_2}$ as reported in Ref.~\onlinecite{piaggi2018predicting}. The free energy of each polymorph was estimated in a range of $(\bar\theta_1, \bar\theta_2)$ centered around the landmarks defined in Table.~\ref{tab:landmark} which allows us to account for thermal fluctuations. It can be seen in Fig. \ref{fig:freeEdiff} that our 1-d RC performs as well if not better in terms of accuracy and convergence speed relative to the 2-d metadynamics.

From simulations biasing $\chi_6$, the free energy difference between liquid state and phase I is 63 $\pm$ 15 $kJ/mol$, while the difference between liquid state and phase IV is and 317 $\pm$ 90 $kJ/mol$. This is qualitatively consistent to experimental observations that form I is the most stable configuration of urea molecules relative to the high pressure product form IV. The other polymorphs which we do not find biasing $\chi_6$ have to the best of our knowledge never been synthesized experimentally, thus reflecting their extremely low probabilities. We further tested the experimental reliability of the obtained Gibbs free energy differences by doing a back-of-the-envelope calculation of the approximate pressure at which urea undergoes a phase transition, and the form IV becomes more stable relative to form I. \cite{Lee2014practical} To do so we assume a linear model $\Delta G(p,T) \approx \Delta G(p_0,T) + (p-p_0)\Delta V$. This gives us the phase transition pressure is approximated to be 1.27 $\pm$ 0.58 $GPa$ for I-IV transition, which is in excellent agreement with the value of 0.69 $GPa$ at 433K reported from experimental measurements. \cite{bridgman1916polymorphism} 

We need to point out configurations sampled in metadynamics simulation performed may have defects such as vacancies or even edge dislocations. These phenomena come from the difference in commensurate ratio in periodic lattice and simulation box.\cite{SMAC2015insight} Seemingly good order parameters may not well represent a state when the underlying structure is distorted, and therefore, affect the extrapolation of thermodynamic and kinetic information. Another challenge is that of finite size effects. The size of simulation box (864 atoms) utilized in this work is much smaller than typical system size, of around $10^4$ Lennard-Jones particles \cite{blow2021seven}, used for study nucleation process, and definitely much smaller than experimentally relevant sizes. Many works\cite{matteo2016cn,salvalaglio2015molecular,Joswiak2018ion,baron2006obtaining,zeiler2012numerical, Sosso2016microscopic, rosales2020seeding, hussain2021howto} have reported strategies for dealing with this which could be useful for making more quantitative comparisons of free energies and eventually kinetics\cite{matteo2016cn,tsai2019reaction} with experiments, which was not the objective of the present work.

\section{Conclusion}
\label{sec:conclusion}
In summary, in this work we have demonstrated how one can learn a 1-dimensional approximate reaction coordinate for urea polymorph nucleation through the SGOOP approach.\cite{tiwary2016spectral}  The RC is expressed as a linear combination of six order parameters which take into account various thermodynamic and structural information. In addition to previously used order parameters such as entropy, enthalpy, coordination number and others, we also introduced averaged angles $(\bar\theta_1, \bar\theta_2)$ as order parameters which can be used to distinguish experimentally observed forms I and IV. Entropy was found to have the highest weight in the reaction coordinate, while coordination number, not surprisingly, had the lowest weight as it can only distinguish liquid from solid but not polymorphs from each other. Biasing along our 1-d RC with well-tempered metadynamics, we could obtain frequent back-and-forth transitions between liquid to those experimentally observed polymorphs within shorter simulation time compared to previous work, and not visiting the polymorphs not reported in experiments. We also performed free energy analysis and showed that form I has lower free energy than form IV. This is also in agreement with the experiment fact where form I was seen in the ambient condition while form IV can only be seen in high pressure condition. Our calculations lead to an accurate estimate of the pressure at which phase transition between these forms happens.  We thus believe that this work represents a step forward towards more automated all-atom resolution sampling and understanding of polymorph nucleation in generic, complex molecular systems.

\section{Acknowledgments}
\label{sec:acknowledgements}
The authors thank Pablo M. Piaggi for providing source codes for the  entropy order parameter and associated WTmetaD setups. We also thank John D. Weeks, Yihang Wang, Dedi Wang, Zachary Smith, and Renjie Zhao for discussions and Luke Evans, Yihang Wang, Dedi Wang for proofreading the manuscript. This research was entirely supported by the U.S. Department of Energy, Office of Science, Basic Energy Sciences, CPIMS Program, under Award DE-SC0021009. We also thank Deepthought2, MARCC and XSEDE (projects CHE180007P and CHE180027P) for computational resources used in this work. Averaged intermolecular angles order parameter code is available at github.com/tiwarylab/ManyAngle-OP.

\end{document}